\documentclass{elsarticle}

\usepackage[utf8]{inputenc}

\usepackage{amssymb}

\usepackage{boxedminipage}

\usepackage{listings}

\usepackage{minitoc}

\usepackage{ifpdf}

\usepackage{graphicx}
\usepackage{paralist}
\usepackage{placeins}
\usepackage{subfigure}
\usepackage{url}

\begin{document}
	
\begin{frontmatter}

\title{Transferring Interactive Search-Based Software Testing to Industry}

\author[1]{Bogdan Marculescu}
\author[1,2]{Robert Feldt}
\author[2]{Richard Torkar}
\author[1]{Simon Poulding}

\address[1]{Blekinge Institute of Technology, Karlskrona, Sweden}
\address[2]{Chalmers and the University of Gothenburg, Gothenburg, Sweden}


\begin{abstract}
	
	\textbf{Context:} Search-Based Software Testing (SBST), and the wider area of Search-Based Software Engineering (SBSE), is the application of optimization algorithms to problems in software testing, and software engineering, respectively. New algorithms, methods, and tools are being developed and validated on benchmark problems. In previous work, we  have also implemented and evaluated Interactive Search-Based Software Testing (ISBST) tool prototypes, with a goal to successfully transfer the technique to industry. 
	
	\textbf{Objective:} While SBST and SBSE solutions are often validated on benchmark problems, there is a need to validate them in an operational setting, and to assess their performance in practice. The present paper discusses the development and deployment of SBST tools for use in industry, and reflects on the transfer of these techniques to industry.
	
	\textbf{Method:} In addition to previous work discussing the development and validation of an ISBST prototype, a new version of the prototype ISBST system was evaluated in the laboratory and in industry. This evaluation is based on an industrial System under Test (SUT) and was carried out with industrial practitioners. The Technology Transfer Model is used as a framework to describe the progression of the development and evaluation of the ISBST system, as it progresses through the first five of its seven steps.
	
	\textbf{Results:} The paper presents a synthesis of previous work developing and evaluating the ISBST prototype, as well as presenting an evaluation, in both academia and industry, of that prototype's latest version. In addition to the evaluation, the paper also discusses the lessons learned from this transfer. 
	
	\textbf{Conclusions:} This paper presents an overview of the development and deployment of the ISBST system in an industrial setting, using the framework of the Technology Transfer Model. We conclude that the ISBST system is capable of evolving useful test cases for that setting, though improvements in the means the system uses to communicate that information to the user are still required. In addition, a set of lessons learned from the project are listed and discussed. Our objective is to help other researchers that wish to validate search-based systems in industry, and provide more information about the benefits and drawbacks of these systems.
	
\end{abstract}


\end{frontmatter}

\pagebreak
\tableofcontents
\pagebreak

\section{Introduction} 
\label{sec:introduction}

	Search-based software testing (SBST) is the application of optimization algorithms to problems in software testing~\cite{McMinn2011, AfzalTF09}, with new algorithms and approaches being proposed and evaluated. Efforts have been made to ensure that these new approaches receive rigorous evaluations, and benchmarks have been developed to enable comparisons between different approaches and their respective evaluations. One example of developing and evaluating new approaches is our work with the Interactive Search-Based Software Testing (ISBST) system. The ISBST system was proposed~\cite{marculescu2012concept}, was evaluated both in academia~\cite{marculescu2015tester} and in industry~\cite{marculescu2014initial}, and further refinements have been proposed~\cite{marculescu2016exploration}. Thus, the ISBST system has been evaluated and validated in academia, and preparations for its transfer to industry are ongoing.

	Successful transfer of SBST to industry would enable companies to improve the quality of their software quality assurance process, with limited resources. In addition to being an effective solution to real engineering problems, successful transfer would also have academic benefits, both in terms of the increase in quality and efficiency that existing evaluations claim for SBST, and in terms of generating additional information, validating existing approaches, and refining our understanding of the underlying phenomena. 
	
	In this paper, we will use the model of technology transfer to industry proposed by Gorschek et al.~\cite{gorschek2006}, henceforth referred to as the Technology Transfer Model or TTM, to evaluate our attempts at transferring SBST to industry, as well as discussing the lessons learned during the transfer process.

	This paper will present our work evaluating and validating the ISBST system. We will use the Technology Transfer Model to assess the maturity of the ISBST system and to frame the lessons learned from its development and evaluation. Section~\ref{sec:related_work} of the paper discusses related work. Section~\ref{sec:method} discusses the context, our industrial partner, and describes the artifacts used in the study. It also presents a synthesis of the development and evaluation of the ISBST system within the framework of the Technology Transfer Model. Section~\ref{sec:static_validation} describes the static validation of the latest version of the ISBST system, on-site, using industrial code and performed by industrial practitioners. Section~\ref{sec:lessons_learned} discusses the lessons learned throughout the development and evaluation of the ISBST system, from its conception and up to, and including, the current study. Section~\ref{sec:threats_to_validity} considers the threats to the validity of the work to develop, assess, and deploy the ISBST system, from its conception until the present version. Section~\ref{sec:discussion} discusses some of the implications of the study, and Section~\ref{sec:conclusions} presents our conclusions.


\section{Related Work} 
\label{sec:related_work}

	Search-based software engineering (SBSE) is an umbrella term coined by Harman and Jones~\cite{Harman2001} to describe the application of search techniques to problems in software engineering. These techniques include both exact, e.g.\ Integer Linear Programming~\cite{integerProgramming}, and metaheuristic techniques, e.g. Differential Evolution~\cite{Storn:1997:DEN:596061.596146}. The branch of SBSE that focuses on software testing is known as search-based software testing (SBST). The application of SBST has been discussed in detail by McMinn~\cite{McMinn2011} for functional, structural, and temporal aspects of testing, and by Afzal et al.~\cite{AfzalTF09} for non-functional testing. 
	
    Efforts to validate SBST with industrial code do exist. Notable is Fraser and Arcuri's EvoSuite~\cite{Fraser2011}, a tool that aims to generate test cases for Java code. The tool has received considerable evaluation,  by Fraser and Arcuri~\cite{Fraser2015} on both open source code and by Campos et al.~\cite{Campos:2014:CTG:2642937.2643002} on industrial code. Doganay et al.~\cite{6571665} conduct an evaluation of a hill climbing algorithm on industrial code derived from Function Block Diagrams developed by their industrial partners. Enoiu et al.~\cite{7515454} conducted an experimental evaluation, also on industrial code, and with master students as experimental subjects. 

	All these evaluations are conducted by researchers on open source or industrial code, and there is little discussion of transferring the tools used to practitioners. Such a transfer, even it its initial stages, has the potential of showing problems that have thus far been ignored and further avenues for improvement. An evaluation by Fraser and Arcuri on the difficulties encountered in applying EvoSuite ``in the real world''~\cite{6569748} discusses the fragility of research prototypes and mentions that even EvoSuite was lacking essential functionality that would allow it to work ``on real code''. That study identifies a number of challenges and classifies them into the Usability (e.g.\ readability of the resulting test cases), Engineering (e.g.\ integrating with the environment), and Research (e.g.\ data collection) categories. 
	
	The assessment of SBST on industrial code is an essential first step towards transferring this technique to industry. In spite of their rigor and depth, however, these studies do not show a complete picture of how SBST could be transferred to industry. The tools developed and presented are often used by researchers and students, rather than industrial practitioners, and the evaluations are conducted on ``historical'' code, rather than living projects that are still in development. The issue of how these findings, tools, and techniques can be transferred to industry is seldom discussed. 
	
	Vos et al.~\cite{Vos2013} also discuss the use of evolutionary techniques for black box testing in an industrial setting. In addition, the transfer of the technique to industry is also actively discussed and considered. The authors conclude that the technique was successful, that evolutionary functional testing is ``both scalable and applicable''. Nevertheless, they concluded that ``a certain level of evolutionary computation skill'' is necessary to allow prospective users to define and refine a suitable fitness function, and that the process of defining the fitness function is time consuming. Thus, transfer to industry would depend on ensuring that prospective users have such skill, or can be supported by researchers. This difficulty in defining a fitness function, together with the need for guidelines and benchmarks, are identified as significant factors preventing more widespread use of evolutionary testing in industry. 
	
	The interaction between search-based systems and their users has also been explored. Users of search based systems can define specifications~\cite{Feldt02}, or interact indirectly~\cite{Feldt99, parmee2000multiobjective}. A more direct type of interaction involves the user directly in the process of assessing solutions that a search-based system finds. For example, Takagi defined Interactive Evolutionary Computation to allow the user to guide a search-based system according to their ``preference, intuition, emotion and psychological aspects''~\cite{Takagi2001}, while Tonella et al.~\cite{Tonella:2010:UIG:1915081.1916189} proposed a system that allowed the user to intervene to break ties in fitness scores. Other approaches involve adapting the fitness calculation to account for user preference~\cite{4804706, 5585740}, to include elegance~\cite{simons2012elegant}, or to ensure that candidates that are known to be good receive a higher fitness score~\cite{liapis2012limitations}. Existing work focuses on interaction with users, but often this interaction is assessed in isolation. In industry, the interaction between the user and an SBST system takes place in the wider context of the organization's software development and testing processes. The exact interaction between the user and a search-based system is contingent on many factors, e.g.\ the intended users, the intended goal of the application, the context. Pannichella et al.\ conclude that ``understandability of test cases is a key factor to optimize in the contest of automated test generation''~\cite{Panichella2016}.
	
	It is also relevant to discuss existing work on the transfer of technology to industry. Gorschek et al.~\cite{gorschek2006} present a technology transfer model that seeks to assess how a research result can move from academia to industry. They describe a number of steps, going from evaluation in academia, static evaluation, and dynamic evaluation in industry. This work provide a useful lens through which the maturity of existing SBST systems can be assessed, and missing elements can be identified.


\section{Context and Artifacts} 
\label{sec:method}

\subsection{The Technology Transfer Model} 
\label{sub:ttm}

The Technology Transfer Model (TTM) proposed by Gorschek et al.~\cite{gorschek2006}, describes seven steps that technology transfer projects go through, along with guidance about putting each of the steps into practice. The TTM steps are:
\begin{enumerate}
    \item Problem Identification. This step focuses on understanding the context of the industrial partner that will be the beneficiary of the technology transfer project. Understanding the domain, establishing a common terminology, understanding and prioritizing the needs of the industrial partner are identified as key issues at this step. 
    \item Formulate a research agenda. Based on the needs identified and prioritized at the previous step, researchers formulate an agenda for their work, in close cooperation with their industry contacts. 
    \item Formulate a candidate solution. A candidate solution is developed for the context, or adapted to fit the context.
    \item Validation in Academia. Once the solution is developed, it is validated in a laboratory setting. 
    \item Static Validation. Static validation consists in having practitioners evaluate the candidate solution, providing feedback to further improve the candidate solution. This type of evaluation takes place in an industrial setting and uses industrial artifacts, but is not carried out in an active project. 
    \item Dynamic Validation. Dynamic validation consists in evaluating the candidate solution as a pilot in industry. This step is aimed at further improving the solution and indicating what is needed for the full scale transfer. The dynamic validation is carried out as part of an active pilot project. 
    \item Release the Solution. This step involves delivery of the candidate solution to industry, along with documentation and reference guides, training support, and measurement programs. 
\end{enumerate}

The model identifies a number of key issues for the successful transfer of technology to industry. First is the matter of identifying the context and understanding the needs of the industrial partner. Second, the importance of adapting a candidate solution to the context, of tailoring the solution to fit the problem and the company. Finally, the model describes an iterative approach to validation, with the candidate solution being validated first in an academic setting, then being subjected to a static validation on historical data, and then a dynamic validation, in active projects. In addition to increasing the realism of each validation, the model argues that additional information emerging from these evaluations could lead to further modifications and improvements to the candidate solution. Thus, each validation step can lead to a re-appraisal of the candidate solution, and can lead to improvements being made. The updated candidate solution is then subjected to the same set of validations, until it is ready for deployment. 

\begin{figure}
	\centering
		\includegraphics[scale=1]{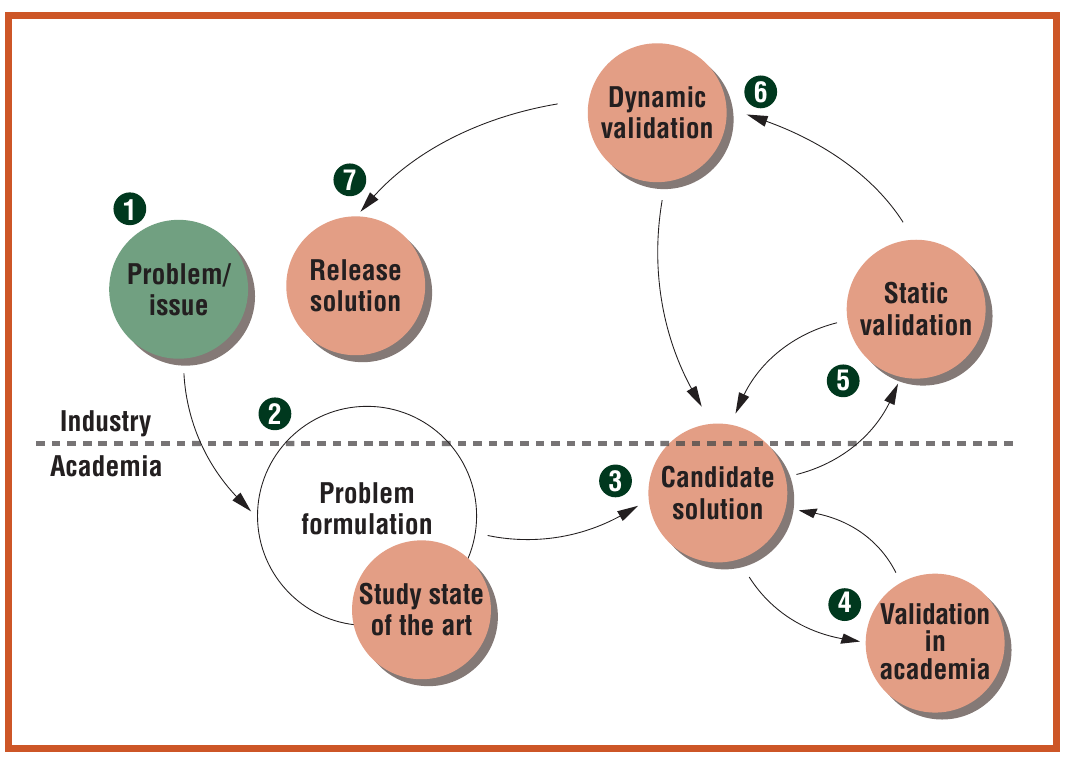}
	\caption{Overview of Technology Transfer Model proposed by Gorschek et al.~\cite{gorschek2006}.}
	\label{fig:ttm}
\end{figure}

The TTM forms a useful framework for discussing the transfer of SBST in an industrial setting. SBST methods have been assessed in academia, and according to the rigors and standards of the academic environment. Using the TTM as a framework allows researchers to extend the assessment to include issues that are important to industry partners as well. 

The ISBST system was developed in collaboration with industry, and went through a number of different versions before the current evaluation. Feedback from our industrial partner was essential in developing the ISBST system in a direction that allowed the updates to be interesting to the company as well. Thus, the ISBST system developed in ways that ensured its relevance to our industrial partner and allowed the researchers to benefit from the feedback of domain specialists.

The ISBST versions cover the first five steps of the TTM, including static validation. Dynamic validation and Release, i.e.\ evaluation of the ISBST system in an active project and turning it over to industry practitioners, are still items of future work. The evolution of the ISBST version we evaluated in industry will be discussed in the following sections.


\subsection{Industrial Context} 
\label{sub:industrial_context}

Our industrial partner is a company offering hardware and software products for off-highway vehicles, as well as components for those products. In addition to developing and testing embedded software themselves, the company offers an embedded software development environment that allows customers to modify embedded software and develop their own modules. Customers use existing modules and components to build function block diagrams (FBD) with the intended functionality. The diagrams are then translated to code, compiled, and deployed on hardware components. 

The context of our industrial partner, and of their customers, places a premium on domain knowledge, rather than knowledge of software development techniques and approaches. It also emphasizes quality of the software and hardware components, but without making software central to the company's business model. A lot of the engineers working there are specialized in their respective domains, with software development being an important, but secondary, part of their work. We will refer to them as ``domain specialists'' rather than software developers, to emphasize this focus. The company wishes to enhance the software development environment to support the domain specialists in developing and running test cases.


\subsection{The ISBST system} 
\label{sub:the_isbst_prototype}

The ISBST tool is a search-based software testing tool that was developed to allow domain specialists to use their knowledge and experience to guide the search. This guidance is achieved by allowing the domain specialist to change the fitness function guiding the search, and then assess the resulting test cases to further improve their definition of the fitness function. The fitness function is composed of a number of criteria, called search objectives, that measure characteristics of the output or input of the SUT\@. The domain specialist guides the search by deciding on the relative importance of these objectives.

\begin{figure}
	\centering
		\includegraphics[scale=0.6]{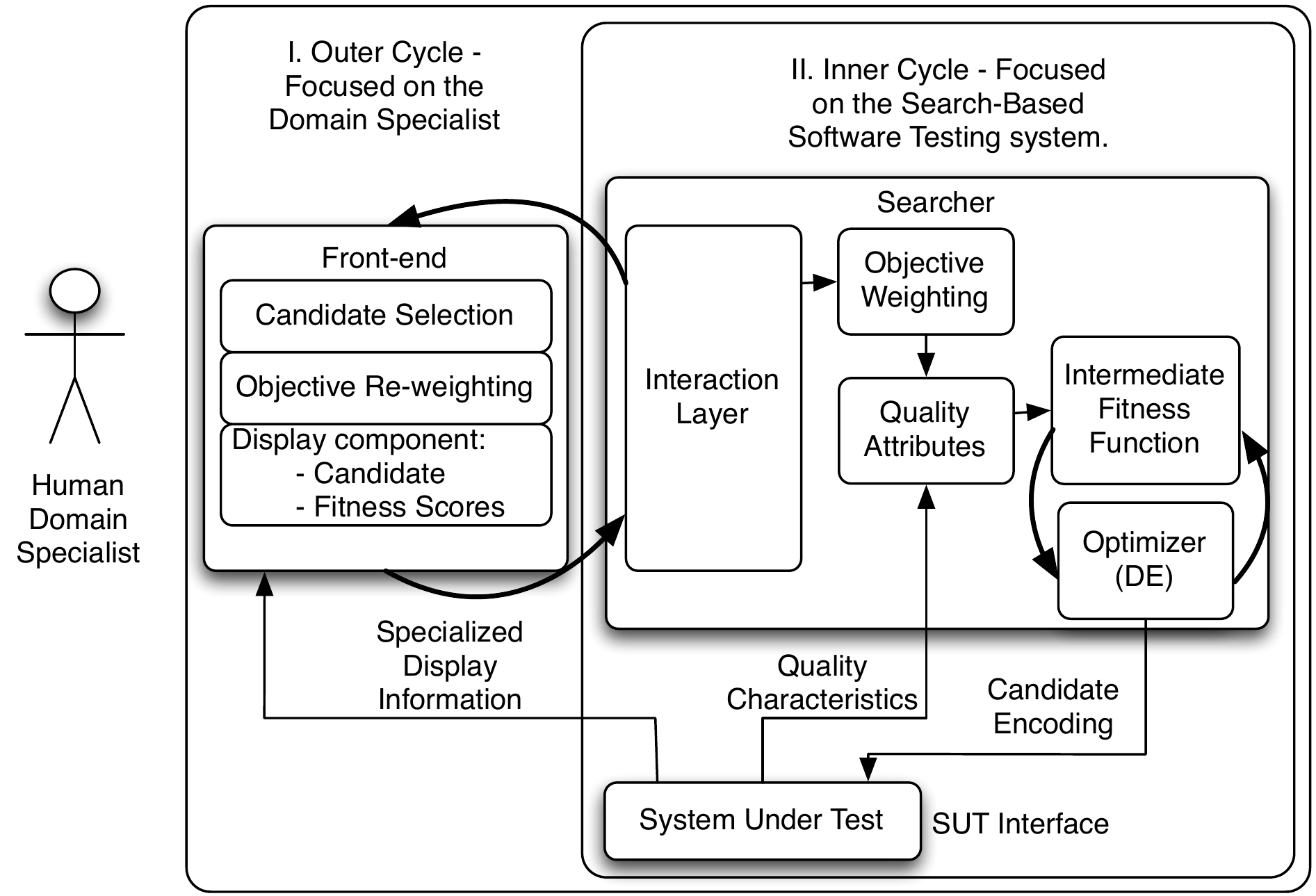}
	\caption{Overview of the ISBST System}
	\label{fig:ISBST}
\end{figure}

The ISBST system has two nested components: an SBST system that connects to the SUT forms the \textit{inner cycle}, and the \textit{outer cycle} that handles the interaction between the inner SBST system and the domain specialist. An overview of the ISBST system can be seen in Figure~\ref{fig:ISBST}.

\textbf{The \emph{Inner Cycle}} consists of the search algorithm itself, the fitness function and the search objectives that form it, and the mechanism that handles the interaction with the SUT\@. The algorithm used is a differential evolution algorithm~\cite{Storn:1997:DEN:596061.596146} that generates a set of $50$ test inputs, that are then used to run the SUT and obtain the corresponding behavior. Each test input consists of a vector of real numbers. The combination of inputs and behavior are referred to collectively as a candidate. Once the behavior has been recorded, the candidate is assessed using the fitness function.

The mutation strategy the ISBST system uses to develop new candidates is as follows:
\begin{equation}
	v_{j,G+1} = x_{r_1,G} + F \times (x_{r_2,G} - x_{r_3,G})
	\label{eq:de}
\end{equation}
where $r_1, r_2, r_3 \in {1, 2, \ldots, NP} $, are integers, and mutually different, and different from the index $j$ of the new candidate. $NP$ is the total number of candidate solutions, and $G$ is the number of the current generation. $F$ is a real and constant factor $\in (0, 2]$ which controls the amplification of the differential variation $(x_{r_2,G} - x_{r_3,G})$. If the mutant vector is an improvement over the target vector, it replaces it in the following generation~\cite{Storn:1997:DEN:596061.596146}.

The crossover rate we used is $cr=0.5$, the scale factor is $F=0.7$, and the population size is $100$. The mutation strategy is that proposed by Storn and Price~\cite{Storn:1997:DEN:596061.596146}: DE\slash rand\slash 1\slash bin. The strategy uses a differential evolution algorithm (DE); the vector to be mutated is randomly chosen (rand); one difference vector is used (1); the crossover scheme is binomial (bin).

The fitness function is made up of several search objectives assessed independently. The results of each of these assessments are collected and combined according to Bentley's Sum of Weighted Global Ratios~\cite{Bentley_findingacceptable}, as can seen below:

\begin{equation}
	\mathrm{DFF}_{j} = \sum_{i=1}^{\mathrm{nObjectives}} { \mathrm{Weight}_{i} * \mathrm{Value}_{i,j} }
	\label{eq:iff}
\end{equation}
where $\mathrm{DFF}_{j}$ (the Dynamic Fitness Function) is the fitness value of candidate $j$, $\mathrm{Weight}_{i}$ is the current weight of the objective $i$, and $\mathrm{Value}_{i,j}$ is the fitness value of candidate $j$ measured by objective $i$. The value of ${DFF}_{j}$ is the sum of the weighted fitness values for all $nObjectives$ objectives. An objective $k$ can be deselected from the computation by having $\mathrm{Weight}_{k}=0$.

\textbf{The \emph{Outer Cycle}} is a shell around the SBST system that allows domain specialists to interact with the SBST by adjusting the relative importance of each search objective and to view the resulting candidates. The candidates resulting from the search are displayed as a group, relative to the fitness values they received. Each individual candidate can be displayed in more detail, if a domain specialist deems it useful. The search interaction is conducted by allowing the domain specialist to set the relative weights for each search objective. The weights are then passed to the Inner Cycle, where they form a part of the fitness evaluation. 

Candidate solutions are displayed, and interaction is permitted after a fixed number of iterations of the \emph{Inner Cycle}. For the system presented and evaluated in this paper, interaction was set to take place every $n_{iterations} = 50$ iterations of the \emph{Inner Cycle}.

At the moment, new search objectives can only be added by hand, with the code for the fitness evaluation being added to the appropriate module. Once the code is written, however, the new search objectives are automatically used for future fitness evaluations. However, experience has shown that any set of search objective that is pre-defined is unlikely to be complete, so a means of allowing new objectives to be added would be useful for practical deployment and further evaluation.


\subsection{The development and previous evaluations of the ISBST system} 
\label{sub:evolution}

In addition to hardware and software, our industrial partner provides their customers with a development environment that allows customers to modify and develop embedded software. The project to transfer SBST to industry was based on the need of our industrial partner to enhance their development environment to also provide support with developing test cases for the embedded modules being developed.

The flexibility of SBST, along with the capabilities exhibited in other domains, make SBST a good candidate for providing testing support for a wide variety of modules. Thus, SBST was chosen as the underlying mechanism for test case generation. The prospective users would be domain specialists, so we decided to encapsulate the SBST component to allow a user to guide the search without requiring them to become specialists in search-based techniques. The first two steps of the TTM, the problem identification and research problem formulation, were carried out iteratively, using the problem formulation to validate and improve our understanding of the research problem, and allowing this improved understanding to refine the research agenda.

The candidate solution we envisioned was an Interactive Search-Based Software Testing (ISBST) system. We decided to develop an interaction component, that would allow the domain specialist to contribute their domain knowledge and experience to the search. Thus, the domain specialists would have an intuitive interface to define the direction of the search without the need to become experts in search-based software testing. An initial design for the ISBST system was proposed~\cite{marculescu2012concept}. In the context of the TTM, the formulation of the candidate solution is defined as a single step, but in practice, the candidate solution is redefined and updated as more information becomes available from the validations in academia and industry. An overview of the latest version of the ISBST system can be seen in Section~\ref{sub:the_isbst_prototype}. 

The validation in academia and static validation in industry proceeded simultaneously, focusing of different aspects of the ISBST system. An initial evaluation of the mechanism chosen for guiding the search was conducted in academia~\cite{marculescu2013reweighting}, and a validation of the visualization component was focused on industry practitioners~\cite{marculescu2013vis}. This information allowed us to update the ISBST system prototype, and conduct a static validation~\cite{marculescu2014initial} of the ISBST system in an industrial setting and with industry practitioners. 

The evaluation in industry validated our choice of interaction mechanism and of the general concept of the ISBST system. As stated in the Technology Transfer Model, the purpose of the static validation is to get feedback and ideas for improvements, validating understanding, and giving feedback to the practitioners involved in the assessment phase in the previous steps. Based on the feedback obtained, the ISBST system was updated to improve performance and accessibility. The updated ISBST system uses the executable modules, rather than the manually instrumented code required by the previous version. This means that new modules can just be plugged in, without any additional effort, and the module being tested is the compiled version that would be deployed on hardware. In addition to improvements to the ISBST system, the evaluation methods used were also reviewed and improved. In particular, we identified potential search strategies used by the industry practitioners, and incorporated those strategies into follow-up evaluations in academia. 

The changes to the ISBST system required us to go through the validation in the laboratory and static validation steps again. The additional validations used lessons learned from previous versions and focused on the effect of interaction on the search process~\cite{marculescu2015tester}, and on investigating the use of exploration to augment the ISBST system~\cite{marculescu2016exploration}. 

These efforts, however, validate the updated ISBST system in the laboratory, in an academic setting. Before moving towards deploying the ISBST system, a second iteration of the static validation step is required. The new static validation would use the results of previous evaluations, in industry and academia, to refine the objectives of the evaluation, in addition to using an updated system.


\subsection{System under Test} 
\label{sub:system_under_test}

For the purpose of this evaluation, the SUT used was a Time Ramp module, part of the standard library of modules provided by our industrial partner. The module is often used as a component in function block diagrams (FBD) that describe other software modules. This function block provides a timed transition from one value to another, with additional features such as signal reset. Input data types must exactly match the types indicated in Table~\ref{tab:table_signals}.

\begin{table}
	\footnotesize
	\centering
	\begin{tabular}{| p{1.4cm} | p{1.2cm} | p{1.2cm} | p{9cm} |}
		\hline
		&Signal&Values&Definition   \\
		\hline																 	
		Input\_0&Loop Tm&U16& Processing time of one program loop (Range: 0 to 65535; Unit: 1ms)    		\\
		\hline
		Input\_1&Reset&BOOL&Sets output equal to Reset Val. (T: Use to preload output at startup to non-zero value. F: Output follows other conditions normally.)    																			\\
		\hline
		Input\_2&ResetVal&S16&The value set as the output when Reset is true. (Range: -32768 to 32767)  	\\
		\hline
		Input\_3&Range&U16&The step size of change in output over DecTm or IncTm (Range: 0 to 65535)    	\\
		\hline
		Input\_4&DecTm&U16& The time it takes for the output to decrease in magnitude from Range to 0 (if DecTm = 0, decreasing ramp function is disabled. If 0 $<$ DecTm $<$ LoopTm, LoopTm is used as DecTm; Range: 0 to 65535)     					\\
		\hline
		Input\_5&IncTm&U16& The time it takes for the output to increase in magnitude from 0 to Range (if IncTm = 0, decreasing ramp function is disabled. If 0 $<$ IncTm $<$ LoopTm, LoopTm is used as IncTm; Range: 0 to 65535)     					\\
		\hline
		Input\_6&Input&S16&Input Signal to be ramped. (Range: -32768 to 32767)    \\
		\hline
		Output\_7&Dec&BOOL&Decreasing. (T: The Output is changed by the whole amount of Decrease Time toward zero) \\
		\hline
		Output\_8&Pasv&BOOL&Passive (T: Output is currently unchanged) \\
		\hline
		Output\_9&Output&S16&Ramped version of the Input signal. Will be equal to ResetVal if Reset is TRUE\@. (Range: -32768 to 32767) \\
		\hline
	\end{tabular}
	\caption{The input and output signals of the SUT used for the evaluation. The variable types are: U16 - unsigned 16-bit integer; S16 - signed 16-bit integer; BOOL - boolean value.}
	\label{tab:table_signals}
\end{table}

For this system, a number of search objectives were developed in collaboration with industry practitioners. The objectives were selected from those relevant for our industrial partner, and refined as a result of feedback from domain specialists in previous evaluations. The search objectives used in this study can be seen in Table~\ref{tab:table_objectives}.

Note that developing the search objectives required domain and software development expertise, but did not involve detailed knowledge of the underlying search-based components of the ISBST system. All the objectives have the same form: they compute a single, scalar, fitness score from the inputs and outputs of the SUT\@. As a result, no additional training is required for domain specialists to develop their own search objectives.

\begin{table}
	\footnotesize
	\centering
	\begin{tabular}{| p{2.5cm} | p{2.1cm} | p{8.2cm} |}
		\hline
		Search Objective & Tag & Definition \\
		\hline																
		Minimize Output Minimum & minimum.min &	The minimum value of the output signals is computed. Smaller values have better fitness. Since Output\_9 is S16, and other output signals are Boolean, this value refers to Output\_9. For multiple output signals, this refers to the minimum value of all signals. A similar objective can be developed for each individual signal. \\
        \hline
		Maximize Output Maximum & maximum.max &	The maximum value of the output signals is computed. Higher values have better fitness. Since Output\_9 is S16, and other output signals are Boolean, this value refers to Output\_9. For multiple output signals, this refers to the maximum value of all signals. A similar objective can be developed for each individual signal.	\\
		\hline
		Output Signal Amplitude & amplitude &	The difference between the minimum value and the maximum value of a given signal. Higher amplitudes have better fitness. The objective refers to Output\_9. In the case of multiple output values, the one with the higher amplitude gives the fitness value. Individual versions of the objective can be developed for each signal. \\
		\hline
		Maximize Output Signal Increase & max.increase & Measures the highest increase in the values between consecutive points of a given output signal. Higher increases give better fitness values. In this example, this refers to Output\_9. For multiple output signals of comparable type, the highest increase found gives the fitness value. Individual versions of this objective can be developed for particular output signals.  	\\
		\hline
		Maximize Output Signal Derivative &	max.derivative &	Calculates the derivative of a given output signal. Higher values of the derivative give better fitness values. In this example, this refers to Output\_9. For multiple output signals of comparable type, the highest increase found gives the fitness value. Individual versions of this objective can be developed for particular output signals.	\\
		\hline
		Minimize Output Signal Mean & min.mean &	Calculates the mean of a given output signal. Lower values of the mean give better fitness values. In this example, this refers to Output\_9. For multiple output signals of comparable type, the lowest mean found gives the fitness value. Individual versions of this objective can be developed for particular output signals.	\\
		\hline
		Maximize Output Signal Decrease&	max.decrease &Measures the highest decrease in the values between consecutive points of a given output signal. Higher decreases give better fitness values. In this example, this refers to Output\_9. For multiple output signals of comparable type, the highest decrease found gives the fitness value. Individual versions of this objective can be developed for particular output signals. \\
		\hline
	\end{tabular}
	\caption{The search objectives and their definition}
	\label{tab:table_objectives}
\end{table}

It is worth mentioning that the module used for the evaluation discussed in this study, like the modules used in previous evaluation of the ISBST system, were already in production at the time of the evaluations and were included in the standard library that our industrial partner and their customers use on a regular basis. As a result, those systems had already undergone rigorous testing and have been used extensively. Therefore, we do not expect that more testing will reveal additional faults in these modules. 

The module was chosen for the study since it is a typical software module for our industrial partner. Its inclusion in the standard library and its use on a regular basis also point to a highly relevant and widely used piece of software.


\subsection{ISBST in use} 
\label{sub:isbst_in_use}

This section will provide a short example illustrating how the ISBST system is meant to help domain specialists develop tests for the SUT they are working on. It also provides examples of the way the ISBST system represents the search results and of the visualizations currently being used to clarify the search progress to the domain specialists.

Let us assume that the ramp module described above has been selected. The domain specialist has a number of manually developed test cases, but needs to develop more to ensure the desired level of trust in the SUT\@.

\begin{figure}
	\centering
		\includegraphics[scale=0.6]{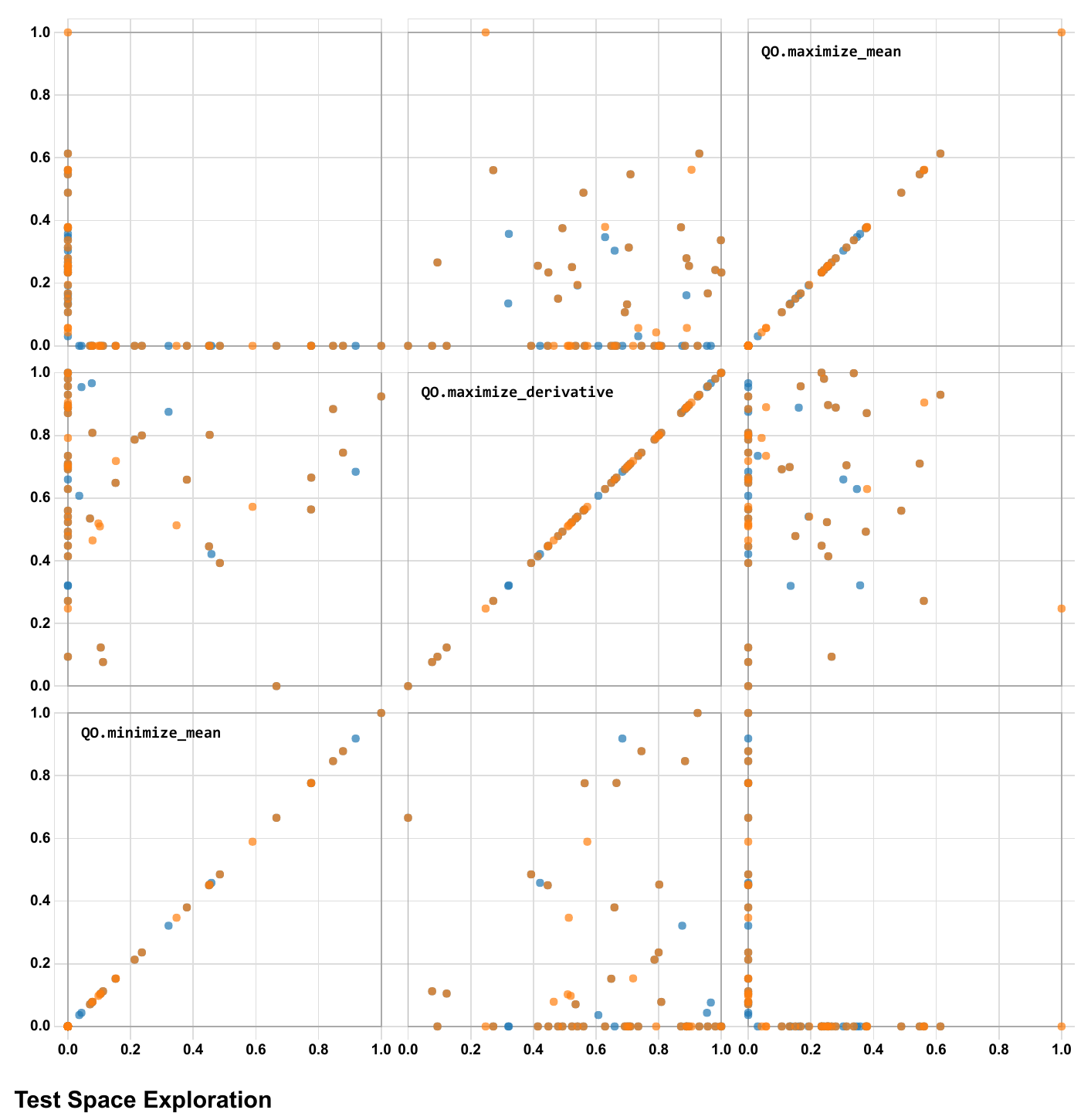}
	\caption{Population of ISBST generated candidates, plotted against 3 of the search objectives. Candidates that are new (i.e.\ from the current interaction step) are blue, those seen in the previous step in orange.}
	\label{fig:tc-exp}
\end{figure}

\begin{figure}
	\centering
		\includegraphics[scale=0.6]{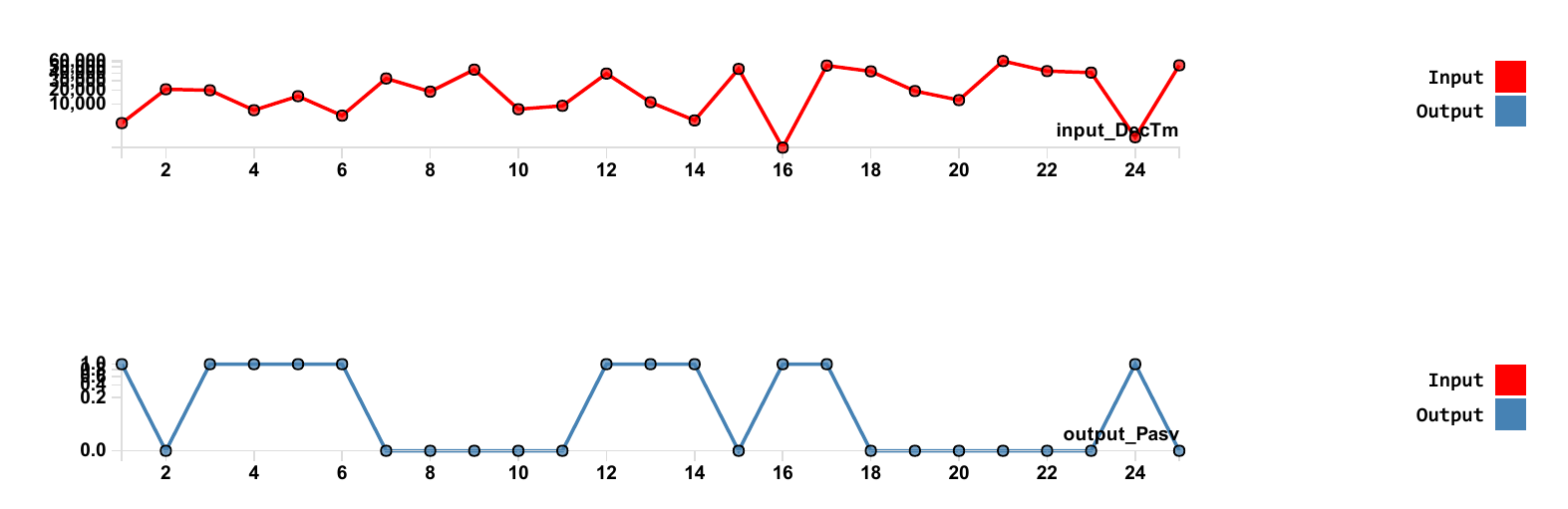}
	\caption{Detailed view of two signals related to one of the test cases. The input signal is in red, the output signal in blue.}
	\label{fig:tc-det}
\end{figure}

The ISBST tool, with the SUT connected, is started and runs an initial number of search steps, with the default values for the search objectives. This allows the ISBST tool to develop a diverse, albeit random, initial population of candidate solutions. The domain specialist chooses the relative weights of the search objectives, and starts the search. After a number of search steps, an \textit{interaction event} is triggered: the ISBST system stops and shows the candidate solutions, as seen in Figure~\ref{fig:tc-exp}, and allows the domain specialist to view the results, adjust the weights of the search objectives, and stop or continue the search. One concern is that of overflows in the internal memory of the module, so the domain specialist selects the ``Maximize Output Signal Derivative" as a top priority. A number of search steps later, the domain specialist sees that the search does not appear to result in further improvements, and selects one of the candidates for closer inspection. An example of this closer inspection panel can be seen in Figure~\ref{fig:tc-det}. The test case in question does result in a sharper than desired change in the output signal, and the matching input values show how that effect can be achieved. The test case is exported for further analysis, and included in future test suites.

With a way to consistently replicate the overflow, the domain specialist can achieve two goals. First, consistently duplicate, and later identify and fix, the problem in the SUT module currently being worked on. Second, ensure that future versions of the SUT module are also tested against similar problems by including the test case in their test suite. This allows domain specialist to generate test cases that have the desired characteristics, while focusing on domain specific tools and skills, rather than on the search-based or software testing domains.



\section{Static validation of the latest ISBST update} 
\label{sec:static_validation}

In the previous section, we discussed the history of evaluation and update that the ISBST system received. As a result of the updates, and in line with the recommendations of the TTM, we ran a second round of laboratory evaluations and validation. The additional evaluations showed that the ISBST system had improved, but further evaluation and validation is still required. This section will describe this round of static validation, highlighting the differences and updates in terms of the ISBST system itself, as well as in terms of the evaluation and validation methods used.

\subsection{Research Questions} 
\label{sub:rqs}

Previous evaluations have focused on the domain specialists' evaluation of the usefulness and usability of the ISBST system, and assessing the effectiveness of the interaction between domain specialists and the ISBST system. As a result of lessons learned in that evaluation, the research questions have been updated. The current research questions make a distinction between the ability of the ISBST system to develop interesting test cases and how clearly the findings of the ISBST system are communicated to the domain specialists.

The study presented in this paper focuses on the following research questions:

\begin{enumerate}
    \item Does the ISBST system develop test cases that can identify bugs in the SUT? We consider that a set of test cases identifies a bug if it causes the SUT versions with the said bug to behave differently from the reference, bug-free, version.
    
    \item To what extent can domain specialists, using the ISBST system, develop test cases that identify the bugs in the SUT? We consider that test cases developed by the domain specialists using the ISBST system identify a bug if that population of test cases causes the SUT versions with bugs to behave differently from the reference SUT version.
    
    \item To what extent does the ISBST system communicate its findings to the domain specialists? Once the ISBST system has developed test cases that can identify a bug in the SUT, can domain specialists clearly identify those test cases as exhibiting interesting or incorrect behaviors? 
\end{enumerate}

The opinions, comments, and feedback of the domain specialists, as well as their subjective assessment of the ISBST system are still of interest, of course. However, the current study focuses more on the ability of the domain specialists to use the ISBST system, to provide guidance for the search that allows the system to develop interesting test cases, and on the ability of the ISBST system to communicate its findings clearly.


\subsection{Method} 
\label{sub:method}

To answer the first research question, a laboratory experiment was conducted. The experiment used the SUT selected and described in Section~\ref{sub:system_under_test}, and the latest updated version of the ISBST system described in Section~\ref{sub:the_isbst_prototype}. The design of the experiment was further improved on the basis of information obtained in previous evaluations regarding the performance and interface of the ISBST system, as well as our improved understanding of the way domain specialists interacted with the ISBST system in previous evaluations. 

The selected SUT is part of a library of modules that have been in use for some time. As a result, the code in question had been thoroughly tested. For the purpose of this validation, we injected $15$ faults, creating $15$ additional SUT versions with bugs to compare against the reference version for a total of $16$ SUT versions. The injected faults were based on existing work that focused on commonly occurring types of faults in this type of system~\cite{Oh2005215, Shin:2012:EEF:2404962.2405003}, with three bugs injected for each category. The exact faults that were injected cannot be discussed in detail, due to the proprietary nature of the code, but the categories of these faults are discussed below.

The categories of faults are: 
\begin{itemize}[1)]
	\item CVR (Constant Value Replacement);
	\item IID (Inverter Insertion or Deletion);
	\item ABR (Arithmetic Block Replacement);
	\item CBR (Comparison Block Replacement);
	\item LBR (Logical Block Replacement).
\end{itemize} 


To reduce the chance that interactions between different bugs would bias the assessment, a separate SUT version was developed for each of the injected bugs, resulting in $16$ different versions of the same system. The ISBST system was used on each of the SUT versions, both with and without the injected bugs, and developed a set of test cases. This set of test cases characterized the behavior of that SUT\@. The behaviors of the bug-injected SUTs were compared against the behavior of the reference, i.e.\ bug-free, original SUT\@.

\textbf{Laboratory experiments.} For the laboratory experiments, the ISBST system was run on each SUT for the same number of interaction events. For each interaction event, the number of fitness evaluations is the same. The number of fitness evaluations is the main metric for evaluating the amount of effort expended by the ISBST system, based on the work of \v{C}repin\v{s}ek et al.~\cite{Crepinsek:2013:EEE:2480741.2480752}. For each SUT the system was run for $10$ interaction events, with $n_{steps} = 50$ optimization steps between interaction events, resulting in a total of $n_{evaluations} = 500$ evaluations of the fitness function for each SUT version. 

We deemed that the bug injected in a particular SUT version was found if the behaviour of that SUT was significantly different from that of the reference, bug-free, versions. The comparison was done based on the search objectives, as well as other metrics, discussed below. The difference was significant if, for at least one of the search objectives, and one of the additional metrics, there was a statistically significant difference between behaviours.

\textbf{On-site evaluation.} To answer the remaining research questions, an on-site evaluation was conducted with three domain specialists from our industrial partner as participants. The evaluation was based on a subset of $6$ SUT versions, the bug free version used as reference, and one version representing each of the injected fault categories. The participants were all domain specialists working for our industrial partner, that had not been directly involved in the development or previous evaluations of the ISBST system.  

The participants were provided with a brief introduction, to familiarize themselves with the ISBST system, the information it provided, and the mechanism for guiding the search. The introduction was a hands-on experience, where the participants ran the ISBST system on the bug-free version. After this introduction, participants evaluated each of the subsequent $5$ SUT versions with injected bugs. Participants were allowed as much time as they needed to complete their assessment, and each participant's evaluation lasted $1-2$ hours. The participants were accompanied by a researcher, to provide answers to questions and to record their feedback and comments. Participants were informed that the $5$ versions had bugs injected, but no additional information was given about the type of bug or the expected change of behavior. 

A lightweight version of the Think Aloud protocol was used to explore the participants' thinking, interpretation of the available data, and to identify any information that is missing, misleading or misinterpreted. The think aloud protocol has been used, for example, for usability engineering~\cite{Nielsen1994TAP}. It involves getting participants to explain their thought process as they go through specific tasks. Nielsen concluded that experimenters need not be highly skilled specialists~\cite{Nielsen1994TAP}, hence the use of a simplified version. In addition, while one participant was not enough to identify usability problems, three or four were found to be enough, with additional participants yielding diminishing returns~\cite{Nielsen1994TAP}. In our study, the goal of the think aloud protocol is to provide a sanity check on assumptions we made about the usability of the ISBST system, and to highlight any usability issues that we might have missed during development. 

\textbf{Assessing behavior differences.} We determine the ISBST system to be successful at finding faults if the population of test cases it produces cause the SUT variants containing faults to behave differently from the bug-free reference version. To determine if a different behavior was observed we use two sets of criteria. The first set of criteria is constituted of the search objectives that are included in the ISBST system and are described in Section~\ref{sub:the_isbst_prototype}. 

In addition to the search objectives, we also developed a number of additional metrics to compare the behaviors of different SUT versions. The additional metrics have been used for subsequent analysis, but were not shown to the domain specialists and did not have an impact on the search process. These metrics can be seen in Table~\ref{tab:additional_metrics}, and have been developed to validate the ISBST system and our previous assumptions:

\begin{itemize}
	\item The objectives that guided the search were developed and selected after discussions with domain specialists, and validated in industry and in academia. Nevertheless, the possibility exists that the behaviors of the SUTs were not completely captured by these objectives. So an additional set of relevant metrics was selected, to further validate the search objectives and provide a better understanding of the SUT behaviors. 
	\item To test the potential for such measurements in future versions of the ISBST system. The current set of search objectives focuses on extreme values in the output signals and on the variation in the output signals. One potential avenue of future improvement for the ISBST system is the development of additional search objectives, using more detailed metrics. One such idea is to measure the distance between input and output signals, and to find test cases where a large discrepancy exists between input variation and output variation. For distance measurements between Boolean signals we used the Longest Common Subsequence, and as a distance measurement between numeric signals we used the Euclidean Distance and the SAX distance~\cite{Lin:2007:ESN:1285960.1285965}. An additional measurement between a current version and a reference population, using Mahalanobis distance, could also be useful for regression testing. 
	\item To illustrate the importance of domain knowledge and SBST knowledge. The measurements compare specific signals based on the assumption that a connection between them is indicative of correct or incorrect behavior. This assumption is based in the detailed knowledge of the particular SUT being tested. Such information is not available to us when developing a general software testing tool, but it is available to the domain specialist, when applying the tool. An example is the Longest Common Subsequence 1-8. The domain knowledge component is that Output\_8 expresses whether the output signal is passive. It shows true in two circumstances: if the previous value of the output signal is equal to the current value, and if the reset signal has been triggered. The SBST knowledge part is that, given the current search algorithm and input value generation, it is unlikely for the input signal to be stable and result in a stable output signal. This would mean that Output\_8 would be true only when the reset signal, i.e.\ Input\_2, is true. 
\end{itemize}

\begin{table}
	\footnotesize
	\centering
	\begin{tabular}{| p{3.1cm} | p{1.6cm} | p{8cm} |}
		\hline
		Additional Metric & Tag & Definition   \\
		\hline																 	
		Longest Common Subsequence 1\-7 & LCS 1\-7 &	 Longest Common subsequence between signals Input\_1 and Output\_7. \\
		\hline																 	
		Longest Common Subsequence 1\-8 & LCS 1\-8 &	 Longest Common subsequence between signals Input\_1 and Output\_8. \\
		\hline
		Euclidean Distance 2\-9 & E 2\-9 & The Euclidean distance between Input\_2 (the reset value signal) and Output\_9 (the output signal). If the reset value is triggered often, the distance between the two signals should decrease. \\
		\hline
		Euclidean Distance 6\-9 & E 6\-9 & The Euclidean distance between Input\_6 (the signal to be ramped) and Output\_9 (the output signal). If the reset value is triggered often, the distance between the two signals should increase. \\
		\hline
		SAX Distance 2\-9 & SAX 2\-9 & The SAX distance between Input\_2 (the reset value signal) and Output\_9 (the output signal). If the reset value is triggered often, the distance between the two signals should decrease. \\
		\hline
		SAX Distance 6\-9 & SAX 6\-9 & The SAX distance between Input\_6 (the signal to be ramped) and Output\_9 (the output signal). If the reset value is triggered often, the distance between the two signals should increase. \\
		\hline
		Mahalanobis distance to reference & M-ref & The Mahalanobis distance from the value of Output\_9 (the output signal) for the current version to the same signal of the reference (i.e.\ bug-free) version. \\
		
		\hline
	\end{tabular}
	\caption{The additional measurements included for the analysis.}
	\label{tab:additional_metrics}
\end{table}

The additional measurements were not presented to any of the domain specialists during the evaluation process, and were applied after the assessments had already been completed. Thus, the additional measurements were only used as an analysis tool. The additional metrics are a diverse set of distances between different signals of the same candidate, or the distance between a certain signal of the candidate compared to the same signal observed in the reference version. A diverse set of distances was used, to ensure a robust evaluation. In addition to the Euclidean distance we also used Symbolic Aggregate approXimation (SAX) Distance~\cite{Lin:2007:ESN:1285960.1285965}, Longest Common Subsequence~\cite{Hirschberg:1975:LSA:360825.360861}, and the Mahalanobis Distance. 

SAX~\cite{Lin:2007:ESN:1285960.1285965} is a symbolic representation of time series that allows a time series of arbitrary length $n$ to be reduced to a string of arbitrary length $w$, typically with $w \ll n$. The algorithm turns a continuous time series in a discrete symbolic representation, that allows the use of existing data-structures and string-manipulation algorithms in computer science. A distance measure can also be defined on this representation. The software developed by our industrial partner and their customers commonly uses time series as input and output signals, so the ability to have a discrete representation for a time series of arbitrary length, as well as a distance defined on that representation, is a useful addition to the set of existing tools. While the input and output signals used in this evaluation are limited to a set number of discrete values, use of SAX as a representation for such signals would allow the distance to be extended to longer input or output signals. 

Longest Common Subsequence~\cite{Hirschberg:1975:LSA:360825.360861} is a way to compare two strings and determine the maximal common subsequence. In our case, domain knowledge provided the impetus for this assessment. For the SUT used in this evaluation, one of the input signals and one of the output signals were known to be equal, under ideal circumstances. While this measure cannot be generalized to other SUTs, it provides a good example of a relatively simple, purpose-build measurement that can highlight obvious faults. When developing the system, we observed that discrepancies between signals that were meant to be identical were easy to identify as faulty, but difficult to observe in the large amount of information being provided and difficult to communicate to prospective users.

Mahalanobis distance~\cite{mahalanobis1936generalised} is a measure of the distance between a point $P$ and a distribution $D$, introduced by P.\ C.\ Mahalanobis in 1936. Mahalanobis distance accounts for covariance between variables, when calculating the distance. In addition, Mahalanobis distance is less sensitive to scale differences between variable values. Thus, variables that are correlated, or that are expressed as higher values, do not unfairly influence the distance measurement. 


\subsection{Results and Analysis} 
\label{sub:results}

\textbf{The laboratory experiment}

We consider that the ISBST system has ``found'' a bug if the behavior observed for the version with the injected bug differs significantly from that of the reference, bug-free, version. Note that this evaluation is focused on the underlying algorithm, and provides little information about the interaction and information communication component of the ISBST system. Assessing how useful or intuitive the interaction is, or how usable the system and how well it integrates with existing tools and processes, could not be done in any meaningful way in academia.

\begin{table}
	\centering
	\begin{tabular}{| p{2.7cm} | p{0.3cm} | p{0.3cm} | p{0.3cm} | p{0.3cm} | p{0.3cm} | p{0.3cm} | p{0.3cm} | p{0.3cm} | p{0.3cm} | p{0.3cm} | p{0.3cm} | p{0.3cm} | p{0.3cm} | p{0.3cm} | p{0.3cm} | p{0.3cm} |}
		\hline
		SUT version&1&2&3&4&5&6&7&8&9&10&11&12&13&14&15&16				\\
		\hline
		minimum.min&&&&&&x&&&&&&&&&&									\\
		maximum.max&&x&&&&&&x&x&x&&&&&&									\\
		amplitude&&&&&&x&&x&x&x&&x&x&&&									\\
		max increase&&&&&&x&&x&x&x&&x&x&&&								\\
		max derivative&&&x&x&&x&&x&x&x&&x&x&x&x&x						\\	
		min mean&&x&&x&x&&&&&&x&&&&&									\\
		max decrease&&&&&&x&&x&&x&&x&x&&&								\\		
		\hline
		LCS 17 &&&&x&x&x&x&&&x&x&x&&&x&x				\\		
		LCS 18 &&&&&&&x&&&&x&&&x&&					\\	
		E 29 &&&&&&x&&x&&&&&&&&							\\	
		E 69 &&x&&x&&x&&x&&x&x&&x&&&						\\		
		SAX 29 &&&&&&&&x&&&&x&&&x&									\\		
		SAX 69 &&x&&x&&x&&x&&x&x&&x&&x&x							\\	
		M-ref &&x&&x&&x&&x&x&x&x&x&&&&	\\
		\hline
	\end{tabular}
	\caption{Objectives that show significant differences between SUT versions with injected bugs and the reference version}
	\label{tab:table_obj_lab}
\end{table}

Table~\ref{tab:table_obj_lab} shows the SUT versions that exhibit significantly different behaviors from the reference version, and the objectives that identify those differences. We define significantly different behaviors to be behaviors for which the scores for at least one of the search objectives show a statistically significant difference from the reference version. Note that no single objective can identify all the behaviors for systems with injected bugs, but that all the bugs are identified by one objective or a combination of objectives.

\begin{figure}
	\centering
		\includegraphics[scale=0.3]{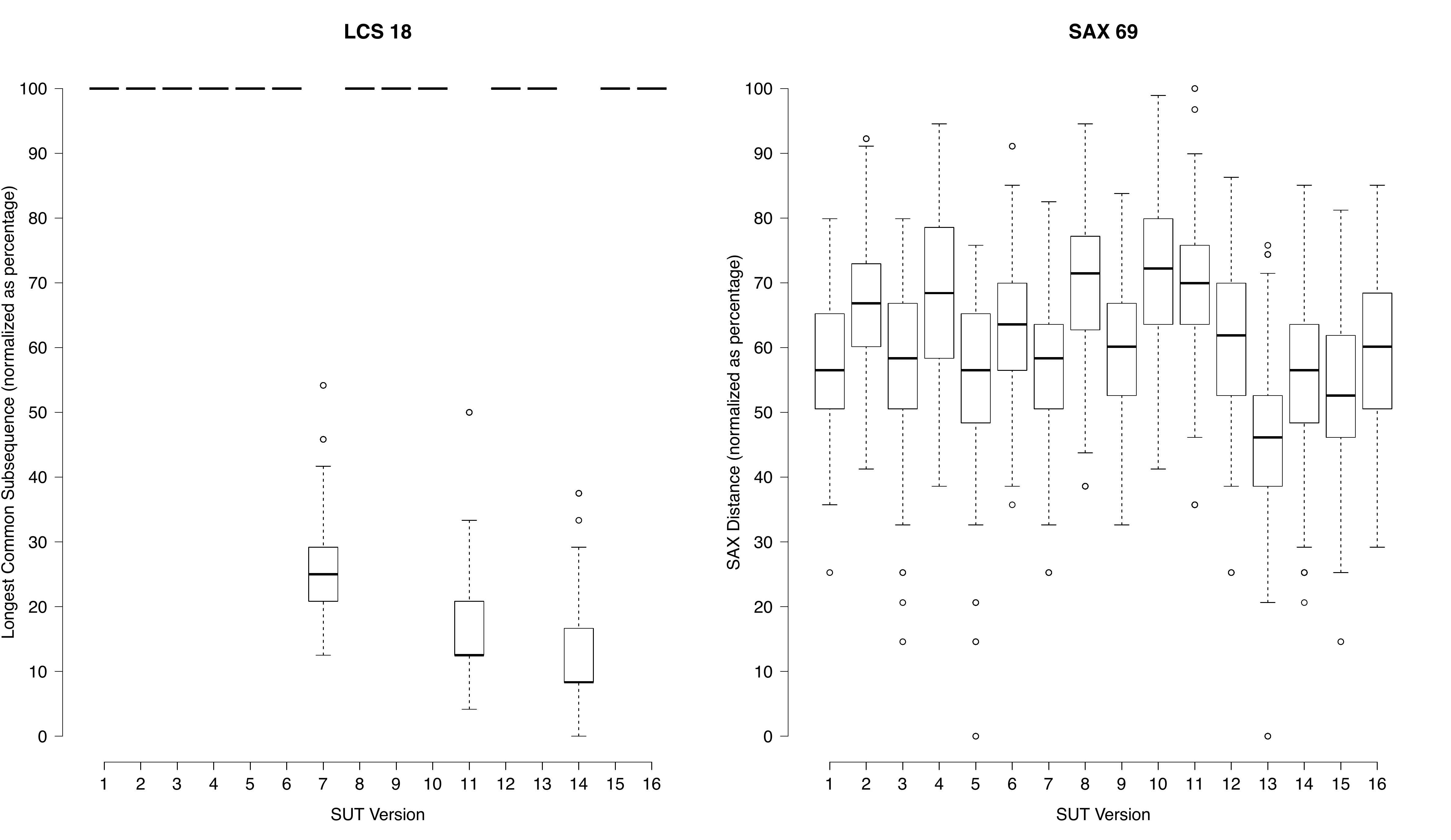}
	\caption{Overview of two of the additional metrics. Longest Common Subsequence Input\_1 and Output\_8 on the left, and the SAX distance between signals Input\_6 and Output\_9 on the right. The X axis shows the respective SUT versions, with 1 being the bug-free reference version. The Y axis shows the values for the respective metrics, normalized as percentages of the minimum and maximum values seen in the dataset.}
	\label{fig:longest_lab}
\end{figure}

Figure~\ref{fig:longest_lab} shows an example of two of the additional metrics that highlight the different behaviors between SUT versions: the Longest Common Subsequence between signals Input\_1 and Output\_8 on the left and the SAX distance between signals Input\_6 and Output\_9 on the right. Knowing the characteristics of the SUT,  signals Input\_1 and Output\_8 should be identical, and any set of longest common subsequence values that is below maximum is ``unexpected behavior'' that should be shown to the domain specialist for evaluation. This is evident for SUT versions $7, 11,$ and $14$. This is one example of how a combination of SBST and domain knowledge can develop a very SUT-specific metric for assessing behaviors. While this metric is not generalizable, similar metrics can be developed for comparing Boolean signals. 

The second metric, the SAX distance, compares Input\_6, the signal to be ramped, to the output signal Output\_9. Distances that are significantly higher than the reference values could mean that the output signal is dissimilar enough from the input not be suitable as a ramped version. Distances that are significantly lower than reference values could mean that the signal changes abruptly, which is what this module seeks to prevent. 

Note that in both examples we rely heavily on domain knowledge to interpret the results of these metrics and draw conclusions from them. 

Table~\ref{tab:table_obj_lab} shows how all the SUT versions with injected bugs show significantly different behaviors, as measured by the objectives we proposed in the ISBST system and by the objectives suggested for later analysis. The differences marked were manually identified as potentially meaningful and found to also be statistically significant. This means that the underlying SBST system is able to propose test cases that cause the SUTs to behave differently. We conclude that the ISBST system is able, under ideal conditions, to identify the injected bugs by comparing the behaviors of the respective SUTs with that of the reference, i.e.\ bug-free, version. This validates the underlying search-based algorithms we used, and increases confidence in the selection of the search objectives that we used, under laboratory conditions. 

On the basis of the results from the laboratory experiment, we can state that the search algorithm, mutation approaches, and selected search objectives were appropriate: the ISBST system was able to distinguish between the behaviors of the buggy versions and that of the reference version of an industrial SUT, under ideal conditions. Thus, in answer to RQ1 we can conclude that the ISBST system does indeed develop test cases that can identify bugs in the SUT, by showing different behaviors between the buggy and the reference versions. 

\textbf{The industrial evaluation}

To assess the ability of the ISBST system to detect the injected bugs, we looked at the behavior differences observed between the reference and the buggy versions of the SUT, by comparing the test case populations developed by each participant for each SUT version. This evaluation is a similar evaluation to that conducted during validation in academia, but applied to the behaviors developed by the domain specialists. Since the validation in academia relied on our interaction model of the domain specialists' interaction with the system, a similar evaluation would show if that model is accurate, or if the assumptions made are correct. The industrial evaluation was carried out with three domain specialists, working as software developers for embedded software with our industrial partner. Their tasks include developing and testing embedded software, as well as providing guidance and support to customers that also develop and test their own embedded systems. The participants were selected from among the engineers that had not participated in evaluation of the previous ISBST system prototypes. 

Due to time and resource limitations, a subset of the bug-injected SUT versions was selected for the industrial evaluation. One version was selected for each of the categories of bugs injected, as mentioned in Section~\ref{sub:method}. In Table~\ref{tab:table_obj_exp}, we define the selected versions as $i\_vj$, where $i$ is the identifier number for the experimental evaluations, and $j$ is the corresponding identifier for the same system in the laboratory experiment. 


\begin{table}
	\centering
	\begin{tabular}{| p{3cm} | p{0.9cm} | p{0.9cm} | p{0.9cm} | p{0.9cm} | p{0.9cm} | p{0.9cm} |}
		\hline
		SUT Version & 1\_v1 & 2\_v4 & 3\_v7 & 4\_v16 & 5\_v11 & 6\_v8\\
		\hline
		minimum.min & & & & & & 1																		\\
		maximum.max&&&&&3&																		\\
		amplitude&&&&&2,3&1,2																		\\
		max increase&&&1&1&1&1,2,3																\\
		max derivative&&1&&1&1&1,2,3															\\
		min mean&&&&1,2&&																		\\
		max decrease&&&&&&																		\\
		\hline																				
		LCS 17 &&&1,2,3&&1,2,3&												\\
		LCS 18 &&&1,2,3&&1,2,3&												\\
		E 29 &&&&&&1,2,3															\\
		E 69 &&1,2,3&&&1,2,3&1,2,3												\\
		SAX 29 &&&&&&1,2,3																	\\
		SAX 69 &&1,2,3&&&1,2,3&1,2,3														\\
		M-ref &&1,2,3&1,2,3&1,2,3&1,2,3&1,2,3							\\
		\hline
	\end{tabular}
	\caption{Objectives that show significant differences between SUT versions with injected bugs and the reference version in the industrial evaluation. The numbers indicate which candidate's data shows significant differences between the injected versions and the reference version.}
	\label{tab:table_obj_exp}
\end{table}

Table~\ref{tab:table_obj_exp} shows the significant differences in behavior as identified by the data resulting from the industrial evaluation. The numbers shown represent the participant that provided data that showed a significant difference in behavior for the respective SUT version with a given objective. All the participants were able to use the ISBST system to generate test cases, and those test cases did exercise behaviors in the SUT variants injected with bugs that differed from the behavior of the reference variant.

However, Table~\ref{tab:table_obj_exp} also shows that different participants exercised different behaviors of the SUT versions. A combination of the initial search objectives (above the line) and the supplementary objectives (below the line) shows that the different versions did indeed show different behaviors, and that the injected bugs did manifest in their behavior. However, some of the SUT versions only showed behavior differences, with respect to the initial search objectives, for one of the participants. This suggests that, for some participants, bug-injected versions were indistinguishable from the bug-free reference version. Additional search objectives show differences, but the initial objectives can be seen to be unsuccessful in identifying different behaviors. 

Note that the behavior of the search objectives with respect to the SUT variants is consistent with the behavior observed in the laboratory. Notice that, not all domain specialists were able to identify differences between the different versions, in spite of using the same SUT versions and the same ISBST system. Given the importance of personal experience, it is not unexpected to see different domain specialists using the ISBST system in different ways. However, since the additional objectives do detect differences between the behaviors observed, and the included search objectives do not, this suggests the objectives are not as robust or reliable as needed, but that improvement can be made to them. 


The overall conclusion is that validating the objectives in an academic setting, under ideal conditions, does not guarantee that those objectives will be as effective in findings bugs in an industrial setting. Different domain specialists may guide search objectives according to different strategies, interpret the findings in different ways, and therefore achieve different outcomes. For example, participant 1 was able to generate test cases that lead to different behaviors for some of the SUT versions, but not for others. The additional objectives performed more consistently, indicating that the SUT versions did indeed result in different behaviors. In practical terms, test cases that point towards a problem could be developed, but not recognized as such by the search objectives. As they would not receive a high fitness value, they may not be shown to the domain specialists, and may therefore be ignored. 

This means that the objectives we have selected for inclusion in the ISBST system are not always good at indicating faulty behavior to the domain specialists, in spite of the previous evaluations in workshops and in laboratory experiments. The differences between expected and observed search objective performance is difficult to estimate without an evaluation in industry. These findings suggest that validation in academia is no substitute for an on-site evaluation carried out with practitioners. In the short term, the search objectives that are already identified as suitable can be included in the ISBST system. For researchers developing search-based systems and seeking to transfer them to industry, this underlines the importance of extensively validating the mechanisms chosen for fitness evaluations in their appropriate context. 


Therefore, in answer to RQ2, we can state that domain specialists, using the ISBST system in an industrial setting, were able to develop test case populations that captured differences between the behaviors of SUT versions with injected bugs and the reference SUT version. This supports the conclusion that the ISBST system is valid in the context we have evaluated it in.

\textbf{Communicating search results}

The issue communicating the results of the search to domain specialists is a much harder problem to assess. For this evaluation, a researcher was present to make note of the interaction between domain specialists and the ISBST system, as well as to collect impressions, comments, and suggestions given by the domain specialists during the evaluation. Overall, we noted that domain specialists were able to quickly adapt to the interaction mechanisms and were able to use them effectively to guide the search. It is worth mentioning, though, that additional explanations and discussion were necessary regarding the operational details of each of the search objectives. 

The information display, however, was more problematic. The general display shows all the current generation of test cases, along with the previous generation of test cases, plotted with respect to their fitness scores. Based on our observations, participants had trouble in discerning what items of the information displayed were relevant. The graphs' axes were pairs of search objectives, with the participant being able to select which objectives to display. However, this resulted in a large amount of information that had to be accessed in separate graphs. The information was available, however the default visualization did not always include relevant information, or the participants' attention was not drawn to relevant candidates. As a result, we observed that some relevant information was not noticed, and participants had a hard time identifying problems in the behaviors. 


For information on individual test cases, candidates that were deemed to have ``interesting'' behaviors, were identified as outliers in the general graphs and selected for visualization. Nevertheless, the characteristics that made those candidates outliers, according to the ISBST system, were not communicated well to the domain specialists. As a result, the domain specialists overlooked those characteristics and were unable to identify those behaviors as either correct or incorrect. 

The current version of the ISBST system includes no definition of what candidate solution can be defined as ``good enough''. Since the number and type of search objectives is arbitrary, it would be difficult to develop threshold values that are meaningful in every situation.

The guidelines we have used in the past for identifying interesting candidate solutions involve looking for extreme values for one or more of the search objectives. The ISBST system also displays both the current and the previous generation of candidate solutions. This enables domain specialists to notice whether or not the search is progressing, by observing the relative difference between generations. 

Based on observations made during the evaluation, the amount of information in the current visualization is somewhat overwhelming. Domain specialists had difficulties in identifying extreme behaviors in the candidate solutions, or extreme values in the search objectives. From this we conclude that future visualizations would need some support. This can range from simple identification of candidate solutions that exhibit extreme values, to more complex techniques of clustering similar candidate solutions, or tracking the progress of the search. More research would be needed to achieve the goals of visualizing search information and search progress, and of communicating this information to the domain specialist in a clear and informative way.

As an answer to RQ3, we conclude that the ISBST system is capable of generating test cases that exercise different behaviors for a given SUT under industrial conditions. However, the mechanisms the system uses to display that information to the domain specialists do not seem able to communicate information clearly enough for practical use. In particular, the ISBST system does not clearly express why certain candidates got the fitness scores they did, so domain specialists have a difficult time assessing their behaviors. Further work is necessary to ensure that the ISBST system communicates its findings in a clear and intuitive way to the domain specialists, and ensures that the reasoning behind the assessment it provides is clear to the domain specialists.


\subsection{Discussion on the transfer of ISBST to industry} 
\label{sub:discussion}

The ISBST system was developed with an industrial partner and is, ultimately, aimed at transfer to industry. The goal for this particular industrial actor is that of enabling domain specialists to develop better quality test cases for their SUTs with less effort. The quality of test cases can be seen as exploring areas of the search space that a human might not think to look at, purposely targeting behaviors that could be faulty or problematic, or validating behaviors that are desirable. By lowering the effort needed for generating test cases, we hope the ISBST system would enable domain specialists to focus on using their domain knowledge and expertise to improve the quality of their work without increasing the cost of testing and test generation.

From the perspective of the Technology Transfer Model, the experimental evaluations presented here fit under two steps. The first, the laboratory experiment, falls under validation in academia. It uses industrial code, but is conducted exclusively in the laboratory. It focuses on the ISBST system itself, on the ability of the system to interact with the SUT and create the appropriate test cases. The second experimental evaluation falls under the static validation step. It is conducted on site, with industrial practitioners as participants, and using an industrial SUT\@. While not part of an active project, the experimental evaluation provides useful feedback about the degree to which the ISBST system could fit in the development environment of our industrial partner.

Previous versions of the ISBST system required source code to be instrumented, to allow test cases to be generated and run. This approach was flawed for two reasons. First, the code being tested did not necessarily behave in the same way as the final product. Since the C code we were testing was further compiled, it could be subjected to optimization that the ISBST system could not account for. There is also the possibility that the instrumentation itself could alter the behavior of the SUT\@. As a result, the updated ISBST system uses the executable file, the version that is ready for deployment on the hardware, with no need for additional instrumentation or manipulation. We can, therefore, argue that the SUT behavior observed in this evaluation is likely to be closer to behavior in use, and less likely to be influenced by our tools. It is worth pointing out that interaction between hardware and software could also alter the behavior. However, once the system is considered stable enough to be deployed on hardware modules, it is subjected to further testing and quality assessment, and the methods and techniques for that stage are already in place. 

The experimental evaluation in industry concluded that further improvements are needed, particularly in the degree to which information is communicated to the domain specialists. Test cases that exhibit extreme behaviors need to be better highlighted, as are the reasons for which those behaviors are considered extreme. It is worth pointing out, however, that the domain specialists adapted quickly to the ISBST system and were able to use it with little intervention from the researcher present. 

Based on our experience, solving technical issues, e.g.\ connecting to the SUT without the need for code instrumentation or manipulation of the artifact, has proven to be more suitable for an academic environment. Potential solutions to this problem could be developed and evaluated in the laboratory, without requiring external resources. Assessing the clarity of the communication between the ISBST system and the user, however, requires the participation of industry practitioners. Potential solutions for this problem need to strike a balance between assessing as many potentially useful methods as possible, and saturating industry practitioners with evaluations, and risk wasting their time.


\subsection{Conclusions} 
\label{sub:conclusions}

As a result of the two experimental evaluations presented above, we conclude that the ISBST system, in its latest iteration, is capable of developing test cases that cause faulty SUTs to exhibit different behavior than the reference, i.e.\ fault-free, versions. As a result, we conclude that the ISBST system has been validated in academia, and is able to develop relevant test cases in ideal conditions.

The experimental evaluation in industry shows that the ISBST system can work under realistic conditions, and that domain specialists are able to use the system to develop test cases that identify faulty behavior. It also showed, however, that further improvements need to be made, in order to allow the ISBST system to clearly and meaningfully communicate its findings to the domain specialist. Clear and meaningful communication of the result findings would enable domain specialists to more accurately guide the search, but would also allow them to better understand how the ISBST system fits in the company's quality assurance process, how it interacts with other tools that support that process, and allow them to provide feedback regarding aspects of the system that need improvement. 

Within the framework of the TTM, the next step towards technology transfer is dynamic evaluation in an active project. While further work is still needed to prepare the ISBST system for transfer, we conclude that the ISBST system is a viable candidate for transfer. 



\section{Lessons Learned} 
\label{sec:lessons_learned}

This section discusses the lessons drawn from the evaluations of the ISBST system presented in this and previous studies, and discusses some of the pitfalls encountered thus far. In the previous and current evaluations, we have gone through $5$ of the $7$ steps of the technology transfer model proposed by Gorschek et al.~\cite{gorschek2006}.

\subsection{Lessons on the transfer of SBST to industry} 
\label{sub:gen_lessons}

In the following, we will discuss lessons on the transfer of an SBST system to industry, that are based on pitfalls that we encountered or narrowly avoided.

\textbf{The need for continuous gathering and validation of information throughout the process.} 

The initial search objective development and selection is based on existing bug databases, and on workshops and interviews with domain specialists. Such initial information may be incomplete, leading to invalid search objectives and approaches, and reducing the relevance of the resulting solution. Thus, we suggest that continuous validation efforts are necessary to ensure that domain knowledge that the objectives are based on is relevant and up to date. The initial efforts to capture relevant information will be incomplete. As more information becomes available, researchers are better able to formulate relevant questions, and domain specialists have a better understanding of what information is necessary, and a clearer set of questions to answer. As a result, we suggest continuously validating the available information, the resulting search objectives, and the selection of search objectives.

\textbf{Search objective selection.}

The selection of search objectives is conducted constantly, with irrelevant objectives being removed, and new objectives being added. 

A search objective is selected if is aimed at defining a set of search objectives that can detect changes in behavior caused by the existence of bugs, and validating that selection. Search objective selection can be a problem if it is based on incomplete information. Categories of bugs that are not present in the initial information, in the bug databases, or are not mentioned by domain specialists, may be missed by the developers. As a result, the selection of objectives may not be able to detect behaviors that are indicative of those types of bugs. 

The search objectives used in this study were selected by researchers in collaboration with the domain specialists. We wanted search objectives that would be interesting from both an academic and a domain perspective. In practice, however, domain specialists would have to look at the search objectives that they find relevant and meaningful. For search-based techniques like the ones we have used, this means objectives that show gradual change as they approach relevant behaviors.

\textbf{Bugs that affect the overall behavior of a system.} 

Certain categories of bugs may change the entire behavior of a particular SUT\@. For example, replacing a constant value that is used in an additive process might change all the outputs consistently. As a result, specific search objectives may have to be developed specifically for that type of bug. A potential solution could be a comparison between the behavior of the current SUT and some reference set, for example resulting from running the previous versions of the same SUT\@.

\textbf{Domain knowledge compromise.}

There are two major forces acting on the researchers when developing the search objectives. The first is a desire to minimize the number of search objectives, and to make them as general as possible. This offers benefits in terms of reuse and in terms of generalizability of the results. The second is a desire to incorporate as much domain knowledge as possible, especially when trying to find specific bugs. A compromise is needed to ensure that the search objectives that are used are both useful for the SUT at hand and generalizable. 

One example of this emerged during our work applying the ISBST system to the TimeRamp module. One input signal is used to transmit a configuration from previous calculations to the present module. Although the interface is defined as U16, only $3$ of the values are meaningful, as they transmit preset information to the module. A SUT-specific search objective could be developed to restrict the search and ensure that computing time, and domain specialist attention, are not wasted on test cases that have meaningless inputs. At the same time, however, this type of search objective can only be used for this SUT\@. Moreover, generating a large number of SUT-specific search objectives may result in problems with the selection of search objectives appropriate for a given SUT\@. 

We suggest that a compromise can be reached, with search objectives that are generalizable being the main core. Where necessary, flexible categories of objective could be developed, that would allow domain specialists to limit the search on a case by case basis. All these efforts, however, would have to be carefully assessed and validated throughout the process.

\textbf{Finding a compromise between robustness and early validation.} 

When evaluating the ISBST system in industry, robustness was a major concern. The system has to be robust enough to be used by domain specialists and should not be prone to random failures or require very specific behaviors from its users. For example, the search process often takes a few seconds. If interaction with the ISBST system during that interval can result in crashes or unpredictable behaviors, this should either be made clear to the user, or interaction should be prevented at that time. 

Problems relating to the robustness of research tools has been mentioned before~\cite{6569748}, with time and resources being cited as possible causes for this problem. Achieving a compromise between early evaluation of a brittle prototype and late evaluation of a more robust version is a problem that can only be solved on a case by case basis. We suggest that the matter be given active consideration. A brittle prototype may fail to provide the necessary information, and may suggest to industry practitioners that the solution is not ready for transfer. A robust, but late, version could result in considerable re-work and wasted effort, as additional information becomes available.

\textbf{Assessing the suitability of search objectives for industrial use.}

Not all search objectives are suitable for use in an industrial environment. For example, this can be due to brittleness, as discussed above. Another example of this is search objectives that require more time to complete: for the ISBST system, evaluations that take 5-7 minutes were deemed to be too long. Domain specialists using the system became disengaged and found it difficult to use the provided functionality. 

Early validation is essential in identifying such search objectives, and in optimizing them to improve execution time, or replacing them with others that offer comparable results. 

\textbf{The effect of correlation between search objectives.}

We discussed earlier that SUT-specific search objectives may be developed to allow the proposed solution to fully use available domain information. Our evaluation of the ISBST system revealed that this could result in search objectives that are not orthogonal. Correlations between search objectives could adversely affect the search, as the correlated objectives are favored by the same type of behavior and offer higher fitness values than objectives that are not correlated. 

Continuous evaluation of the search objective selection would allow researchers to determine if the search is affected by such behaviors, and to correct any problems with the search parameters. 

\textbf{Tool reliability.}

At this stage, the proposed system is meant to be evaluated in an active project. For this to work, the system has to be reliable enough to use without the constant presence of the researchers and without constant tinkering. At this moment it should be a functional, robust, usable tool. The trade-off we discussed previously, between reliability and early evaluation, stops being an issue. At this stage, the system should have received a significant amount of evaluation and improvement, so resources can be spent on reliability. 

\textbf{The effect of tool usability on the evaluation.}

In our experience thus far, usability has not been considered a priority. Previous evaluations have been mostly academic, using researchers or automated tools to run the system. For the evaluations that we conducted in industry, a researcher was always present to answer questions, provide information and clarification, and fix any problems with the ISBST tool. Low usability could have a negative impact on tool evaluation, as the efforts and feedback of participants focus more on identifying problems with the tool rather than on assessing the underlying concept, or the potential uses and problems with the technique. In any tool that should be evaluated in an active project, and later transferred to industry, usability is worth the resource and time investment.


\subsection{Lessons specific to interactive systems} 
\label{sub:isbst_lessons}

The lessons above are useful to the transfer of SBST systems to industry in a more general sense. The current and previous studies have also revealed lessons that are applicable in particular to the transfer of interactive SBST systems. 

\textbf{Information overload.}

As stated before, a search-based system can generate large amounts of information. The ISBST system displayed a total of $100$ test case candidates, comprising the current and previous populations. They could be displayed relative to each other, in a set of 2-dimensional graphs, one for every combination of two search objectives that the domain specialist wanted to visualize. Each selected candidate could also be visualized separately, with the input and output signals displayed on demand. 

While all this information was useful, not all of it was equally relevant. For example, identifying outliers with respect to individual search objectives was relatively easy, but outliers with respect to several objectives was not as clear. The relatively large amount of information, as well as difficulty in identifying quickly which items of information were more important or relevant, lead to confusion. Domain specialists were lost in the information provided. We suggest that this can happen even for systems that do not require user interaction, but that do involve people in assessing and interpreting the results. 

Early validation could help researchers in identifying this problem. Solutions could vary on a case by case basis. Information could be divided between several different areas of concern, e.g.\ separating the overall view from the display of individual candidates. Visual aides could be provided: outliers or candidates that the system regards as remarkable in some way could be highlighted, and the reason for this selection provided. Not all of these require significant changes to the functionality of the system, but could make an important difference for any efforts at evaluating, validating, and transferring such systems to industry. 

\textbf{Awareness of the search progress.}

Our experience with the ISBST system shows the importance of keeping the user informed of the progress of the search. This will allow users to decide if additional effort spent searching could lead to better results, or if a new approach should be tried. 

For example, the system could show how the overall fitness values have changed during the search, which search objectives have seen improvements in the fitness scores and which have not. This could be useful in determining if the search is going in a desirable direction. 


\subsection{Overall Lessons} 
\label{sub:overall_lessons}

In general, we would like to highlight the importance of early and continuous validation of any tool being transferred to industry. While we assume that any such tools have already been evaluated in academia, they would have to be changed to adapt to the new context, and to fit with the company's tools and processes. Continuous validation ensures that everything from the search algorithm to the interaction and information display mechanisms are appropriate for the task. We also strongly advise that such evaluations are as close to real operation as possible. This means involving practitioners early, considering tool and process interactions, and validating information display mechanisms. Realistic evaluations also allow practitioners at the company to become familiar with the new tools, allowing them to provide more relevant information and to conduct a better and more informed assessment.

A second general recommendation is to keep tool design flexible. From the initial step, where a search-based solution is proposed and validated in academia, until the final step, when it is ready for deployment, a prototype will undergo significant changes. 

Last, we found that communicating information to the industry practitioners is a non-trivial problem. Sufficient information needs to be available to allow practitioners to make an informed decision. That information needs to be presented in a clear and reasonable way, to allow them to quickly understand it and use it for decision making. Domain specific visualization approaches are important, since they are already familiar to potential users and require no additional training to use. 



\section{Threats to validity} 
\label{sec:threats_to_validity}

This section will discuss the threats to the validity of the entire study, from the initial development of the ISBST system, to the most recent evaluation.

The study is based on our experiences developing and evaluating the ISBST system in industry. The development of the ISBST system was focused on the needs of our industrial partner, and on assessing that system in academia and industry. While we made every effort to ensure that our conclusions are accurate, some threats to validity still exist, and will be discussed in this section.

We would like to underline that the list of potential problems is not complete or exhaustive. Further efforts will surely reveal additional problems, and ways of addressing them, especially in the dynamic validation phase. The list of lessons presented here should help researchers in initiating projects to transfer search-based technology to industry and in navigating the early phases of such projects. 

During our study, the first three steps of the Technology Transfer Model overlapped to some extent. As a result, we will discuss the validity threats relevant for those steps together. The ISBST system was developed based on information from, and to address the needs of, a specific company working with a specific type of embedded software. While we have not yet identified any reason why our conclusion cannot be applied for other types of software systems, or why our lessons are not relevant for other contexts, we cannot safely generalize on the basis of this example alone. 

Moreover, the initial information collected from our industrial partner shaped the development of both the ISBST system and of the evaluation mechanisms. Our assumptions were based on the accuracy and completeness of this initial information. In later stages of the project we have made efforts to validate those assumptions, and correct them when they were unsuitable. Nevertheless, it is possible that some of our assumptions are not accurate or generalizable. While we have confidence in our approach and our conclusions, we advise other authors seeking to transfer their research to industry to pay careful attention to their data collection and data validation steps, especially in the initial stages of technology transfer. 

We focused on our industrial partner, their context and tool chain, and on the systems they wanted to test and the problems that they were expecting. As a result, the ISBST tool was designed to fit that context and fulfill those requirements. We have made the ISBST tool flexible: it can use different search objectives~\cite{marculescu2015tester, marculescu2014initial}, and different search algorithms~\cite{marculescu2016exploration}. Nevertheless, it is difficult to know how easy it is to transfer the ISBST tool to other domains until such a project is undertaken and the results analyzed. 

The following steps, Validation in Academia and Static Validation, will also be discussed together. For these steps, we emphasized the importance of the evaluation mechanisms we developed for the purpose, and on the potential differences between evaluations suitable for academic settings, and those suitable for industrial settings. All the evaluations were developed by researchers. Some of the evaluations, particularly in the Static Validation step, were conducted by domain specialists, but under the supervision of researchers, and with data collection being conducted by researchers. While we have made efforts to ensure that the evaluations are as objective as possible, there is a possibility that our own biases have influenced the conclusions. As a result, we encourage other researchers to assess our results, conduct similar projects, and share their findings. To this end, we have prepared a replication package, containing the ISBST system, the evaluation mechanisms, and the analysis scripts we used~\footnote{https://sites.google.com/view/bogdan-marculescu/introduction/additional-material}. 

An essential issue regarding validation is that of visualization. A good visualization is essential to ensure that participants understand the system, its capabilities, and can provide meaningful feedback. For the ISBST system, we developed visualization tools based on the commonly accepted approaches used by the domain specialists, and we continuously improved them throughout the project. The description of an interaction between the domain specialist and the ISBST system can be seen Subsection~\ref{sub:isbst_in_use} and the visualizations in Figures~\ref{fig:tc-exp} and~\ref{fig:tc-det} While we seem to have achieved our goal to make visualization clear enough to allow good evaluations, we have also identified problems. Further work is needed to fully understand how to develop and evaluation visualization mechanisms, especially visualization mechanisms applicable to a more general type of domain.

The static evaluation described above was conducted in an industrial setting, with industrial practitioners and industrial code. That evaluation has provided evidence that the ISBST system we developed is usable, and that search-based software testing is useful in improving the testing process. However, the evaluation was not conducted in a live project. It was also comparatively short. A dynamic evaluation, conducted in an active project, where the ISBST system is used by domain specialists without researcher involvement, and conducted over a longer time span is needed to confirm the usability and usefulness of search-based software testing in an industrial setting and to provide further feedback. 

Lastly, our evaluation was based on a small number of engineers at the company. While only three engineers participated in our study, they do form a significant proportion of the domain specialists working in that business unit. This may limit the generalizability of our conclusions, especially on subjective considerations like interaction evaluation. The participants in this study were engineers at the company, working with the type of SUT that we evaluated on a daily basis, developing and testing similar systems. We argue that, in spite of their low number, their experience and knowledge makes their evaluation useful and meaningful. Nevertheless, further research is needed before a definitive conclusion can be reached.


\section{Discussion} 
\label{sec:discussion}

The study presented above is based on our experiences developing and implementing an ISBST solution for an industrial context. We used the Technology Transfer Model proposed by Gorschek et al.~\cite{gorschek2006} as a framework to assess the progress of the project to transfer the ISBST to industry.

Existing work argues that search-based techniques cannot be used by domain specialists  without the support of experts in evolutionary computation~\cite{Vos2013}. We acknowledge the difficulties in developing a complex and sophisticated fitness function without experience. By using a set of domain specific search objectives, the ISBST system allows domain specialists to set priorities and guide the search without the need to develop a fitness function by hand. Our evaluations in industry show that this approach is intuitive enough to allow domain specialists to guide the search and develop test cases, even in the absence of experience with search-based systems. The study by Vos et al.~\cite{Vos2013} also finds, however, that evolutionary computation is useful in industry and that the results compensate for the time and effort spent. The authors also identify a number of obstacles that stand in the way of transfer of such techniques to industry. Perfecting the way the ISBST system in particular, and other automated test systems in general, communicate their findings to the domain specialist could be crucial in the adoption of such tools to industry. Before showing the usefulness of the automated tools, researchers must ensure that their findings are understandable to the domain specialists~\cite{Panichella2016}. We argue that the ISBST system partially addresses the need for constant support for domain specialists.

We firmly believe that SBSE in general, and SBST in particular, are useful and flexible tools and could provide benefits to industry. The goal of this paper is to promote more applied research into the development, implementation, and application of search-based software systems. More validations in industry would yield additional information about such tools and strengthen confidence in their usefulness. 


\section{Conclusions} 
\label{sec:conclusions}

Search-based software engineering has received considerable attention from researchers. A lot of the research in SBST is focused on developing new search-based techniques, and evaluating and validating them. Tools such as EvoSuite~\cite{Fraser2011, 6569748} provide support for SBST research and have received extensive validation on open-source and industrial systems.

In this paper, we presented a project to develop and transfer an Interactive Search-Based Software Testing system to industry. The lessons learned from our own development and evaluation of the ISBST system should prove useful for developing, deploying, and validating such search-based software tools for use in an industrial context. We encourage researchers to seek early and continuous interactions with industry, to assess and validate their ideas, and to ensure that their efforts are relevant and useful for industrial practitioners. We also discuss the importance of tailoring systems for the benefit of the companies where they should be used. Interaction is a central concept to ISBST, and may be less so to other SBST tools. Nevertheless, visualizing the results of a search in an intuitive and meaningful way, and ensuring that an SBST tool integrates well with the processes of the company where it will be used are essential for the success of any technology transfer.


\section{Acknowledgements} 
\label{sec:acknowledgements}

This research has been supported by funding from The Knowledge Foundation (KKS) in the project  Testing of Critical System Characteristics (TOCSYC), project number $2013/0085$.


\section{References}

\bibliographystyle{elsarticle-num}
\bibliography{p7}

\end{document}